\documentclass[11pt]{article}
\usepackage{epsf}
\usepackage{graphicx}        

\def\c{\cite}

\def\r1{(\ref{$1})}
\def\re{\ref}

\def\ll{\label}

\def\th{\theta}
\def\ba{\begin{array}{c}}

\def\ea{\end{array}}

\def\si{\sigma}

\def\de{\delta}

\def\l{\left}
\def\l({\left(}
\def\r){\right)}
\def\r{\right}

\def\la{\lambda}
\def\al{\alpha}

\def\be{\begin{equation}}
\def\bc{\begin{center}}
\def\ec{\end{center}}
\def\bit{\begin{itemize}}
\def\eit{\end{itemize}}
\def\ee{\end{equation}}
\def\ed{\end{document}}
\def\bea{\begin{eqnarray}}
\def\eea{\end{eqnarray}}
\def\efr{\end{flushright}}

\begin{document}
\title{{ 
Shape changing  and accelerating solitons in 
integrable variable mass sine-Gordon model
}}
\author{ Anjan Kundu\\ 
 Theory Group, 
Saha Institute of Nuclear Physics\\
 Calcutta, INDIA\\
{anjan.kundu@saha.ac.in}}
\maketitle
\begin{abstract}
 Sine-Gordon  model with  variable mass (VMSG)   
  appears in many physical  
systems, ranging from  the  current through nonuniform  Josephson junction 
 to DNA-promoter dynamics.
 Such models are usually  nonintegrable with solutions found numerically or
 peturbatively. We construct a class of VMSG models, integrable both
at classical and quantum level with   exact soliton solutions, which can 
  accelerate,   change their shape, width and amplitude
 simulating realistic inhomogeneous systems at certain limits.
 \end{abstract}

\medskip

Sine-Gordon (SG) model
enjoys a special status among 
 nonlinear integrable systems 
 for  its inherent
 richness and
 wide range of applications in different fields
\c{JJferroNLop,sflux04,sgDNA,msgDNA,dipti,Saction,vsg,ssg07,prl07}.
Apart from the fascinating  properties of      
 an integrable system, 
 the SG model
exhibits  relativistic invariance and
 integer-valued topological charge  represented by  
solutions like
kink, antikink, breather etc. \c{solit1}, together with  
the quantum integrability  described  by
 the 
Yang-Baxter  equation (YBE),
which  for the SG   leads to  
    quantum algebra   $
su_q(2) $ \c{qsg,fadalg}.

  Solitons in the standard SG model, as  in  other 
integrable systems,  move with  constant velocity and  shape. In 
 realistic situations however due to
  inhomogeneity  of the media solitons may exhibit more complex motion with
changing
 velocity  and shape   
 \c{msgDNA,Saction,vsg}, which 
 may be used also  as  a desirable effect
 for 
  fast transport, fast communication,
or even for a possible  soliton gun \c{ssg07}.
In particular inhomogeneity   can  lead to  
 SG models with 
 variable mass  (VMSG) 
 in describing fluxon or semi-fluxon  dynamics in  Josephson junction (JJ)
 with impurity or nonuniform critical current \c{sflux04,Saction},
  spin wave propagation with variable  
 interaction strength
\c{dipti},  DNA-promoter dynamics in  
 nonuniform background
  etc.
However  such  inhomogeneities
  usually destroy the  most cherishable
property   of  the SG  model, i.e.
 its integrability   
 and  hence the   solutions 
 can be extracted 
only numerically or  at
 best peturbatively
\c{sflux04,msgDNA,Saction,vsg}.

 We   observe that, though the  integrability   of
  SG model is  spoiled
 by a variable mass, or  a
 variable  velocity,     it    can be 
  restored if  both of them vary simultaneously following certain
rule. Therefore
we can   construct a VMSG model,  
  integrable 
at the classical and the  quantum level, allowing  
   analytic soliton solutions. Such exact solitons 
 nevertheless show 
intriguing accelerated motion
 with changing   shape, amplitude and width, simulating
 thus realistic inhomogeneous systems 
 \c{msgDNA,Saction,vsg} and describing them  analytically 
at certain limits (Fig. 1b-d).

To clarify  our  strategy 
we focus  on the   linear   spectral problem of the SG model:  
 $ \ \Phi_x(x,\lambda)=U(\lambda, x)\Phi(x,\lambda), \\  
\Phi_t(x,\lambda)=V(\lambda, x)\Phi(x,\lambda), \
$ 
with its Lax pair \c{SG-Lax}:
$ \ U = { \frac i 4} \left( -u_t \si ^3 +m k_1 \cos {\frac u 2} \si ^2
-m k_0 \sin {\frac u 2} \si ^1 \right),
\ \ $ and $
 V={ \frac i 4} \left( -u_x \si ^3 -m k_0  \cos {\frac u 2} \si ^2
+m k_1 \sin {\frac u 2} \si ^1 \right), \ $
where $ k_0(\la) =  2 \la +\frac {1} {2 \la}, \ \ 
k_1(\la) =  2 \la -\frac {1} {2 \la}, $ with
   spectral parameter $ \ \la  $.
Compatibility condition $\Phi_{xt}=
\Phi_{tx} $ or the related  flatness condition
leads to 
the  SG equation,  for
 constant    mass  $m_0 $  and spectral parameter
 $\la _0 $. Recall 
that in the 
 inverse scattering (IS) method  solitons are obtained
at discrete spectrum $\la _n, \ n=1,2, \ldots N $,
with   velocities  of   SG solitons   linked 
 to  these values of the spectral parameter.
Therefore  variable
  soliton velocity   should
have a variable spectral parameter  $\la $, which alone  however 
  violates  the compatibility
condition.  Fortunately  by making 
  mass   $m$ also   a space-time dependent   variable,  
 we can  get the
  VMSG  equation
\be u_{tt}-u_{xx}+m^2(x,t) { \sin} u=0,\ \ 
 , \ll{msge}\ee  
with the
 constraint:
$ (k_{0}m)_t+(m k_{1})_x=0, \  (k_{1}m)_t+(m   k_{0})_x=0 $,
which  can be reduced to two  free field equations:
\be \kappa_{tt}-\kappa_{xx} =0, \ \rho_{tt} -\rho_{xx}
=0, \ \mbox{for} \ \kappa=\ln m(x.t), \ \rho=\ln \lambda (x,t). \ll{ml}\ee
Note that the  set of equations (\re{msge}-\re{ml}) is  a new
integrable relativistic system, generalizing SG equation,
 and a reduction (at the free field limit of spectral dilatation field  $\rho
$) of the conformal
affine Toda  model (CATM) \c{CFTSG}.
 However since we are interested here in  application to  
physically relevant  inhomogeneous models, we would  consider  $\kappa, \rho $
 as given inhomogeneous functions  by  restricting to particular   
  solutions for  variable mass
 and  spectral parameter:
\be m(x,t)=m_0 f_+f_-, \ \ \lambda (x,t)=\lambda_0  \frac { f_+} {f_-} ,
    \ll{mla} \ee
with
 $f_\pm $  
arbitrary smooth 
  functions of $x_\pm=x\pm t $, respectively.
Thus we  obtain an  integrable SG equation (\re{msge}) with  variable mass
$  m(x,t)= m_0 f_+f_- $. Note  that due to explicit space-time
dependent coefficient, it       
   is no longer    a  relativistic and translational 
 invariant model. However if we demand such invariance,  we    
 simply get back the constant mass    SG model, as shown in  \c{Barash87}.
 We can 
control      mass $m(x,t) $ in the  VMSG model 
(\ref{msge}) by     
choosing suitably the inhomogeneity functions $f_{\pm }$ 
 for different   physical situations, showing  a variety of   
     soliton dynamics as in  Fig. 1 a-d. 

For  obtaining  exact
 solutions of the VMSG model 
(\ref{msge}), we can
 apply    well known methods
  designed for  integrable systems \c{solit1}, e.g.
   Hirota's  bilinearization 
 and the 
 IS formalism.
Hirota's    method  for soliton solution for the standard  
SG equation 
is given  through a cleverly chosen ansatz 
$ u=-2i \ln \frac {g_+} {g_-} , $
 with   $ g_\pm $ as conjugate 
functions,
which  converts the SG equation   into  a bilinear
form,   admiting  solution for $g_\pm $
  as expansion   in plane-waves.
For SG  model (\ref{msge})  with
 variable mass  and velocity
 the
 same  ansatz seems to work,  only  the 
plane waves
should be  replaced    by 
their  { generalized } form: 
$\  g^{(n)}=\frac {c_n}{\la _n}e^{\frac i 2 
 (X(\la _n,x,t)-T(\la _n,x,t))}, $ where
\ $X(\la _n,x,t)=\int^x dx'm(x',t) k_{1}(\la _n,x',t), \  T(\la _n,x,t)=
\int^t dt' m(x,t')k_{0}(\la _n,x,t') .$
 This gives the exact
  soliton solutions of  (\ref{msge}) through the expansion:
$ \ g_\pm =1\pm g^{(1)}, \ \ $  { for 1-kink } 
and $ \ g_{\pm} =1\pm (g^{(1)} +g^{(2)})
+s(\frac 1 2 ({\theta_1-\theta _2}) )g^{(1)}g^{(2)}, \ 
\la _a=\frac i 2 e^{\theta _a}, a=1,2
 \ \ $ { for  2-kink }
etc., with the  scattering amplitude $s(\th )= tanh ^2 \th  $, while for
 $ \la _2=-\la _1^*=\eta  e^{i\theta }$, one gets  the  kink-antikink 
bound state (breather solution).

Similarly we can apply the IS  formalism  \c{solit1} to (\ref{msge}), for which
the  crucial step is to  use 
the analyticity   of 
 Jost    solutions $ \Phi $,
 based on their behavior at 
$ \la \to \pm \infty$. This  holds equally 
  for the inhomogeneous  extension of the SG model,  
  where  the asymptotic plane waves 
 should again be  replaced  by their generalized form.
  $1$-kink soliton   with $\la _1 =i \eta $ 
can be obtained   explicitly,
 either from the  Hirota's or 
 from the IS method  as \be  
\ u=4 \tan ^{-1} (e^ {  \zeta }), \ \zeta=
  \frac i 2 (X(i\eta,x,t)-T(i\eta ,x,t)), 
 \ \ll {sinu2} \ee
with  variable soliton velocity   
 $v_s(x,t)=-\frac {dx} {dt}=\frac { k_1(\eta, x.t)} { k_0(\eta,x.t)}$.
Kink solution (\ref{sinu2}) gives a localized soliton for  
$\sin \frac  u 2 =
 \frac {1} {cosh  \zeta }, \ \ $ which are actually shown in  Fig 1.

To see the effect of various inhomogeneities on the  
  soliton dynamics, we consider some concrete integrable cases.
Notice that  
  variable mass i)
 $m_0(x^2-t^2)^{n} $, 
 invariant under relativistic motion,  yields ( for 
 $n=1$)
  the  exact kink solution 
(\ref{sinu2})  with
 $\zeta= \frac m 3 (2 \eta (x-t)^3+\frac 1 {2 \eta }
(x+t)^3) $, the evolution of the corresponding  soliton   is 
 depicted in Fig 1a. The intriguing 
  change in the soliton  shape, width and  velocity 
 during  its motion is clearly seen. 
 Position-dependent mass  can be achieved in this case 
  at    $ t
\to 0$ and therefore
 phenomenon like fluxon propagation through JJ with local
 defect  $m_0x^2 $
may be described with  the above analytic soliton solution
at a short time interval limit,
 as shown in Fig 1b.

   Other forms of  integrable VMSG equations  can be  obtained for 
  mass ii)  $\sqrt 2m_0\cos^\alpha q(x \pm t), $  $ \alpha $ being an arbitrary
parameter.
 We derive for the first case (with  $\alpha =1 $) 
  kink solution  (\ref{sinu2})
 with $
\zeta=  m_0  (k_0(\eta) x-k_1(\eta)t+\frac 1 {4q \eta} \sin 2q(x+t)), \
$
having soliton  velocity 
$ v_s= m_0d (k_1(\eta) - \frac 1 {2\eta } \cos 2q(x+t))$ and   width 
$ d  = 
(m_0(k_0(\eta) + \frac 1 { 2\eta } \cos 2q(x+t)))^{-1},$ both of which
 oscillate periodically in space-time,
as  evident from  Fig. 1c.
Notice that  variable mass of this type is particularly
interesting, since it can  describe an important
physical situation, namely the SG equation parametrically driven
by a plane wave \c{BarashEL91}.

One can  get a similar integrable case with 
 mass iii)
$m_0(2\cos q(x+t) \cos q(x-t))^{\frac  \alpha 2} $,
which  
at short time interval limit ($ t \to 0$) gives
$ \approx m(x)= \tilde m_0(\cos qx)^{ \alpha }$,
 while for evolution limited to a small 
space interval ($x \to 0 $):
 $ \approx m(t)= \tilde m_0(\cos qt)^{ \alpha }$.
 Recall  that
 a physically motivated   spin chain  with 
coupling constant   changing  periodically in space, can be described 
 by a VMSG 
  with  mass  $m(x) =m_0(\cos qx)^\alpha ,$ where $\alpha=\frac 1 {2-K} $,
with $K\geq \frac 1 2  $ being an important parameter of the system  \c{dipti}.
Similarly a real oscillator chain  pumped by an alternating current 
\c{tsg}  can be linked to a VMSG with 
 mass  $m(t) =(\cos qt)^{\frac 1 2} $.  
Therefore we may conclude that the analytic  solution of  our VMSG
equation
can describe the  
 inhomogeneous spin wave dynamics  \c{dipti} or evolution of 
forced oscillators 
 \c{tsg}, 
 at least at  short time or space interval limit.
Alternatively, 
this realistic spin (or oscillator) model 
 can be tuned  to the integrable VMSG with mass $m_0\cos^\alpha q(x + t)$, 
  by making  its
coupling strength oscillate 
     periodically also  in time (or space).

In most physical situations 
the inhomogeneity of the media
 leads to  VMSG equation,
with only  position-dependent  mass 
 $m(x)$.
Therefore we explore to find, when such  equations can be  
 integrable in the entire space-time  and 
 conclude from our result (\ref{mla}),
that the  
 VMSG    becomes integrable only for the   space-dependent  mass iv)
 $m(x)=m_0  e^{\rho \ (x-x_0)}$ with
 $\rho = const. $, 
  which explains    also
why most of the realistic VMSG models  
 having  a different   position-dependent mass (e.g.  \c{dipti})  
turn out to be  nonintegrable systems.
Exact kink  solution for the  integrable $m(x) $  is obtained
from  
(\ref{sinu2}) as $u=4 \tan ^{-1} (e^ {  \zeta }), \ \zeta=\frac 1 \rho k_0(t)m(x) ,
 \ m(x)=   \exp (\rho (x-x_0))  ,
 \ k_0(t)=\cosh (\th -\rho (t-t_0)). $
The corresponding soliton 
velocity and  width  are
 $v_s=tanh (\th -\rho (t-t_0)), $ and 
 $d=( {m(x) k_0(t)})^{-1} $,  showing how the soliton shape   
   changes and  how
it   
    accelerates,  
decelerates or can   exhibit    
 boomeron \c{ColDeg} like property.
 This scenario is  close to the
predicted  behavior of solitons 
  in the dynamically active  promoter zone of the T7A$_1 $ 
DNA 
 \c{msgDNA}.
Notice   that  for $\rho >0,$   since 
$ m(\infty)=
\infty,  \ m(-\infty)=
0  $, the 
kink solution  yields $u(\infty)=2\pi,\ u(-\infty)=\pi $ , and hence 
 corresponds to the  
 topological charge 
 $Q=\frac 1 {2 \pi}(u(\infty)-u(-\infty))=\frac 1 2$. This 
 intriguing fact might serve as an 
 integrable theory for the   semi-fluxon, observed in unconventional JJ \c{sfluxE}. 

At $ \rho \to 0$: $\zeta
\to 
\zeta_0=m_0( k_0 \ (x-x_0)- k_1 \ (t-t_0))$ and 
 the standard SG soliton   is recovered. Therefore  we can 
acess the solitonic behavior 
 in realistic models
with any 
mass deviation from its constant value, 
with  high  degree of accuracy,
by approximating through 
expansion in powers of $\rho $.
Fig. 1d shows that a
 static soliton  remains static, when placed in a region with
 constant mass while an initially static soliton
 can move with accelerated (or decelerated) motion, 
when placed  in a 
  zone with variable mass, resembling  closely the scenario  
 of 
the VMSG   soliton in the DNA chain, which
 with zero initial velocity 
 in an
  inactive regions (with  constant mass
 due to almost uniform background of 
   two types of  base pairs)
   remains static, while
in the  active promoter region with variable mass  
(due to significant difference 
in the number of lighter (A-T) and heavier (G-C) base pairs)
the same   static soliton  can
  acquire rich accelerated motion
 \c{msgDNA}.

\begin{figure}[t]
\epsfxsize=0.96\textwidth
\includegraphics[width=4.cm,height=3.1 cm]{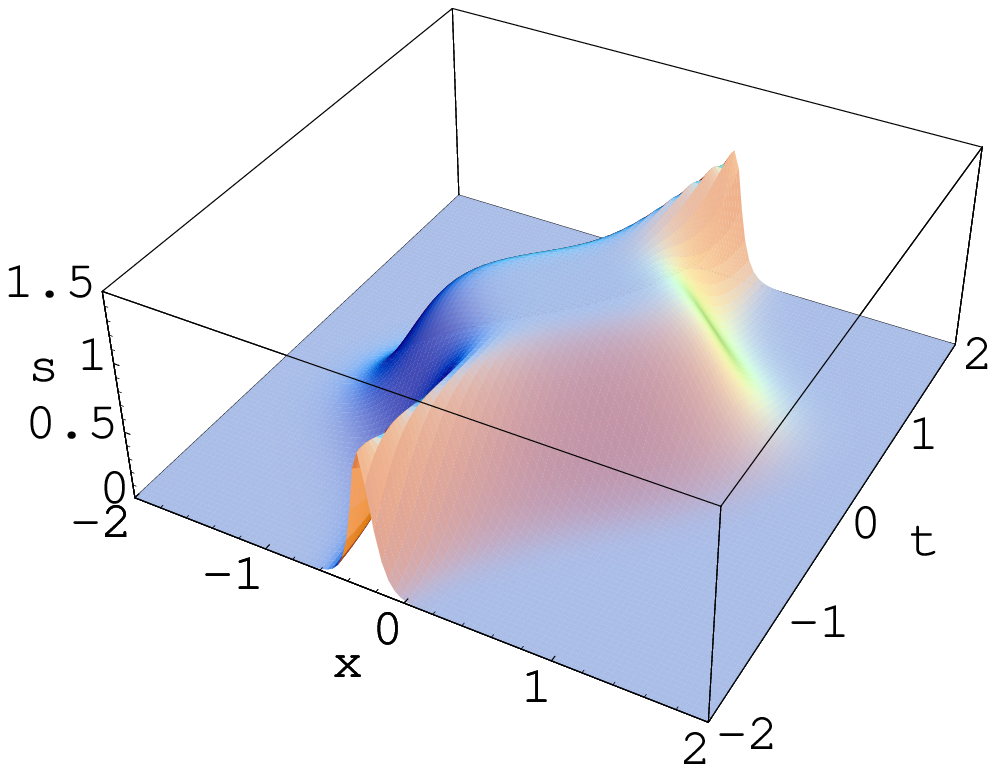}  
\includegraphics[width=4.cm]{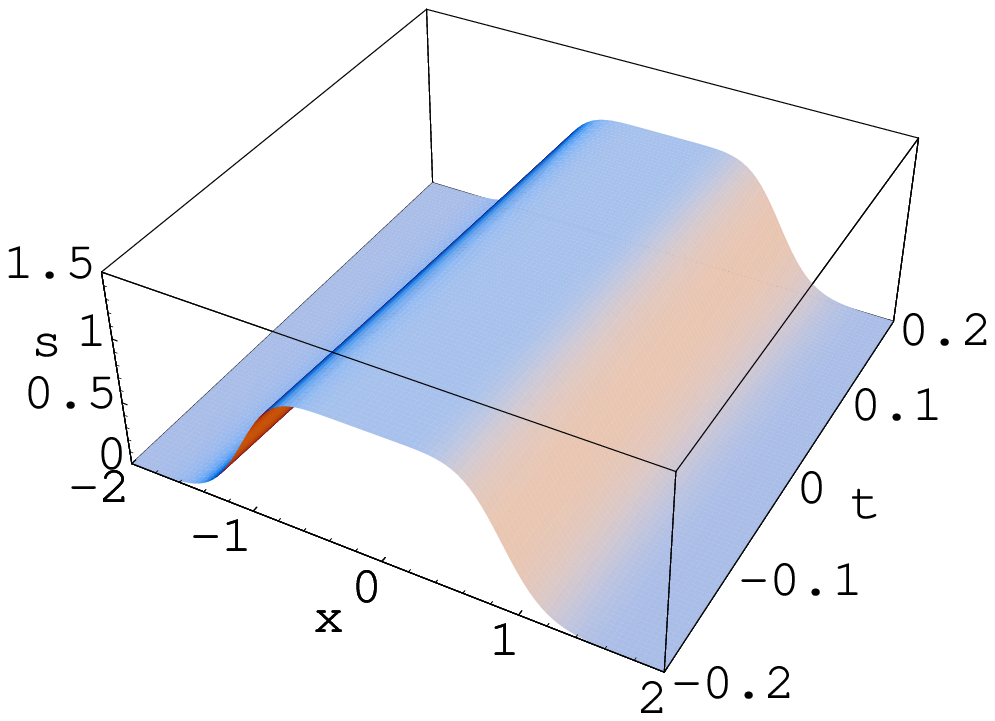} \ \ 
\includegraphics[width=4.cm]{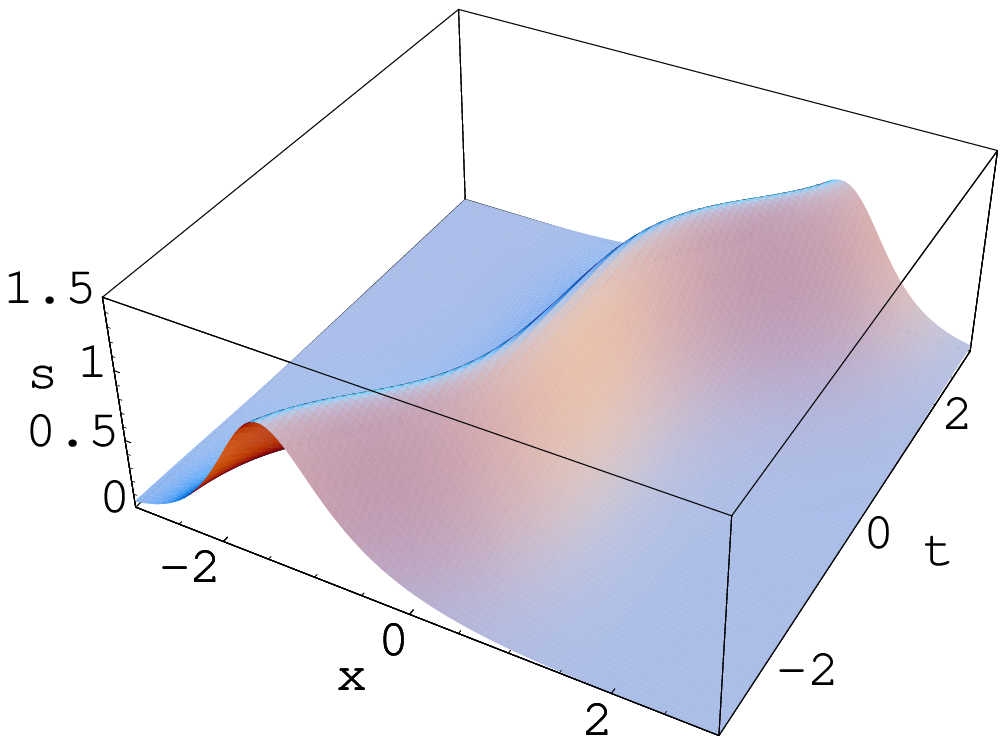} 
\includegraphics[width=4.cm]{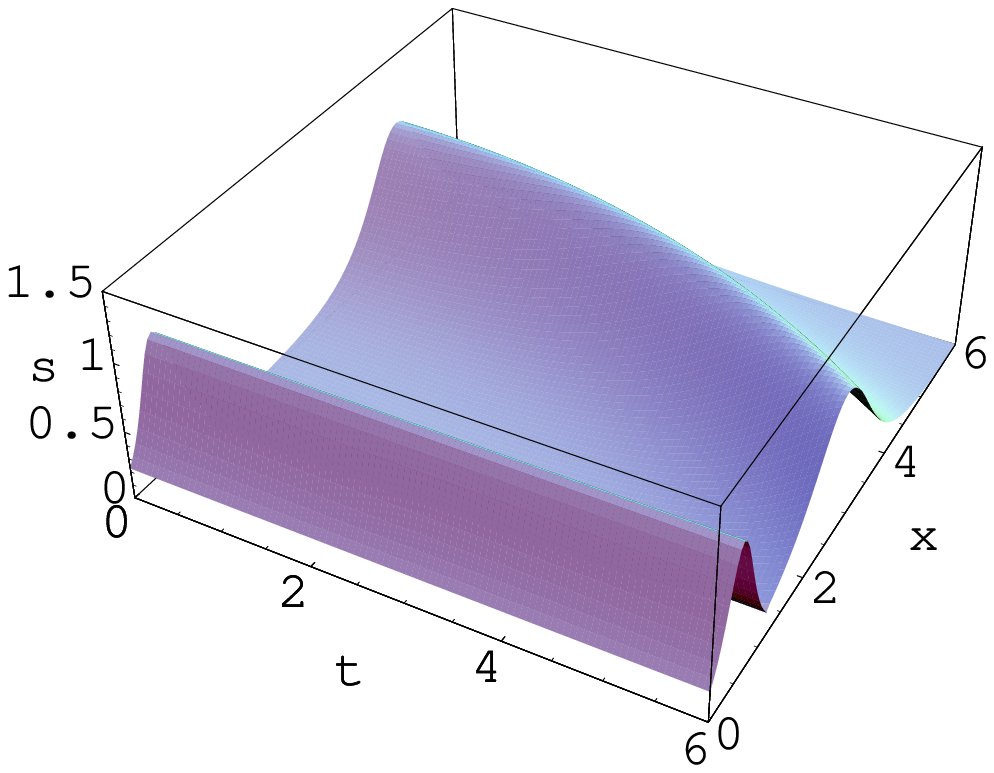}
\caption{   Exact  soliton  solutions: $s=\sin \frac u 2 (x,t)  $
of  the integrable VMSG equation with  variable mass m.  a)
 $m=m_0(x^2-t^2)$ having an intriguing flattening of the soliton 
 at the center. \ b)
Short time interval limit of the above soliton showing  the
flattening prominently. \ c)  $m= 
2m_0\cos q (x+t)$ with  oscillatory  behavior of the  soliton. 
\ d) Static soliton in the zone ($x \leq 1.2 $)
 with
  $m=const $ and initially static soliton   
in the zone (at $x=4.8$)
with variable mass: 
 $m=m_0 \exp (\rho x) \approx m(1+
\rho x)$, with $\rho=0.1 $
 moves backward
 with acceleration,
 resembling
  soliton propagation in 
  inactive/active  promoter region in a DNA chain.}
\end{figure}
Finally we intend to show that the
 integrable VMSG model constructed
 here can be raised to the quantum level 
and  the  algebraic Bethe ansatz  (ABA) developed 
for 
 the constant mass SG model  \c{qsg}
can be adopted successfully for it.
  Quantum lattice regularized SG  matrix Lax operator 
$U_j(\la , {\bf S}_j, m), \ j=1,2,
\ldots , L $
 involves  quantum-spin operators $S_j^3 (u_j),
  S_j^\pm (u_j,p_j,m)$, expressed in canonical 
operators $ \ u_j, \ p_j= \dot u_j \ $ and   
the mass parameter $ m$, which should be generalized now 
to  site  dependent  parameter
$m_j$ \c{prl99}.
Note  that,
 the  trigonometric  $R(\frac {\la } {\mu })$-matrix associated 
with our quantum integrable VMSG model  remains
the same,   
  since it depends  
 on the ratio of two spectral
parameters, 
in which   $x,t $-dependence enters, as seen from   (\ref{mla}),   only   
 multiplicatively  and therefore gets canceled.
Moreover, YBE being a local algebra (at each lattice site $j$) 
is  not affected by  
inhomogeneity and yields    
   the same quantum algebra  $su_q(2) $; only with 
the replacement of  constant
  $m$ by a site-dependent function  $  m_j $ 
 in its structure constant:
$ \ [S_j^+, S_k^-]= \delta _{jk}
 m_j \frac {\sin \al 2S_j^3}
 {\sin \al}.$

The aim of ABA
 is to solve exactly the eigenvalue problem of   
$tr T(\la )=A (\la )+D (\la ), \ $ with $ \ T(\la )=\prod_j U_j (\la ) $,
 generating all conserved
operators including the Hamiltonian, with the eigenstates:
  $ |\la _1, \ldots, \la _n>= \prod_j^n B(\la _j)|0>$.
 $T_{12}(\la) =B(\la )$ acts as  {creation} operator, while
$T_{21}(\la ) =C(\la )$  as  { destruction} operator annihilating
the pseudovaccum: $C(\la
)|0>=0. $
 Following closely   \c{qsg},   
but  generalizing    for
  the site-dependent   mass  $m_j $, we can construct  the 
local  pseudovaccum
 $|0>=\prod_j|\Omega_j^{(2)}> $, 
  a crucial step in ABA, by combining the action of  consecutive 
pair of Lax operators: $U_jU_{j+1}|0> $    \c{fg}.   
Consequently the vacuum eigenvalues are
 generalized for the 
quantum VMSG model as
$A(\la )|0>=\alpha_{(m)}|0> , \ \ D(\la )|0>=\beta _{(m)} |0>$, where
  $\alpha_{(m)}=\prod_j a(\th ,\frac {m_{j}  } 
{m_{j+1} }) , \ \beta _{(m)} =
 \prod_j a^*(\th ,\frac {m_{j+1}  } 
{m_{j} }) $  with $a(\th ,\frac {m_{1}  } 
{m_{2} }) =
  \frac {m_1  } {m_{2} }+ \de ^2 {m_{1}  }{m_{2} }
(\cosh (2 \th +i \al)) $. This   yields the exact eigenvalue for the conserved quantities:
$tr T(\la )$ as \ 
$\Lambda (\la ;\la _1, \ldots, \la n)=\alpha _{(m)}
 \prod_j^n f(\frac {\la _j}{ \la
})+ \beta _{(m)}
\prod_j^n f(\frac {\la } { \la _j}) , \ \ $ 
where $f(\frac {\la }{ \mu
}) $ is expressed through the elements of the   
  $ R(\frac {\la }{ \mu
})$-matrix, which 
 remains unchanged  for the VMSG model. The Bethe equations
 for determining the parameters $\{ \la _j \}$
are generalized similarly by taking $m \to m_j $.

 Since our VMSG  model can be reduced from CATM,  
 a coordinate  
 transformation
 $(x,t) \to (X,T) $ exists, which 
can take the SG model with variable mass   formally  to the SG
  model with constant mass, though  the  domain
  might shift to  an unphysical region and singularities
might arise.  Such a nonlinear   transformation,  amounting to going  
 to a noninertial frame,
 takes particularly simple form in the light-cone coordinates
: $X_\pm=\int   {dx}_\pm f^2_\pm $ \c{Barash87,CFTSG}. 
 However 
for  investigating   
 the physical effect of a given inhomogeneous medium 
inducing  accelerated  and shape changing solitons,
  one has to
analyze the model in  its original form with variable mass.
 Similar situation arises
also in other  inhomogeneous  systems with integrable
  nonisospectral flow, e.g. \c{inlsprl}
in the study of  accelerated solitons in plasma 
through 
  NLS  equation,
 in 
 discrete NLS, 
 in  Toda chain,
 in
   matrix  Schr\"odinger problem with 
 boomeron solution   \c{ColDeg} etc.
In most of these models
 though the inhomogeneities could be removed by tricky
nonlinear transformations, the investigations 
 were carried out in the
original  systems due to their  physical relevance.
Surprisingly  this  long list of literature 
devoted to  various  inhomogeneous integrable 
models   does not include 
the well known SG model and also avoids any quantum treatment,
except  perhaps a  recent attempt \c{kuninhom}.  This   
enhances therefore the
 importance of the present result, which explores the inhomogeneous
integrable  SG model, presents
 its exact solution  both in the  
classical and in the quantum case. Its analytic soliton solutions 
can simulate at certain limits 
  realistic events like, 
  fluxons in nonuniform Josephson junction, 
 dynamics of spin waves with variable coupling, DNA solitons in the active
promoter region etc.
 Regulating the
position-dependent mass in integrable VMSG model one can create semi-kink  solution suggesting a
possible exact theory
 for semi-fluxon.

The author likes to thank Profs. P. Mitra (SINP) and D. Sen (IISC) for
helpful discussions.


\begin{thebibliography}{99}
\bibitem{fg}$| \Omega_j^{(2)}>= 
(1+\de ^2
  g_{m_{2j}m_{2j+1}}(q_{2j},q_{2j+1})) f_{m_{2j}m_{2j+1}}(q_{2j},q_{2j+1})
 $, where $ f_{m_1m_{2}}= \left( {m_2 ^{\frac {q_2} {\alpha}} }/ 
{m_1 ^{\frac {q_1} {\alpha}} } \right) f_m,
 \ g_{m_1m_{2}}=(\frac {m_2} {m})^2 g_m
 $, are  generalizations 
 over their known solutions  $f_m,g_m $ with mass
 $m= const. $ \c{qsg}
\end{thebibliography}
   \end{document}